\documentclass{elsart}

\usepackage{graphics}
\usepackage{graphicx}
\usepackage{epsfig}

\usepackage{amssymb}

\begin{document}

\begin{frontmatter}



\title{NA60 results on the $\rho$ spectral function in In-In collisions}


\vspace*{-0.2cm}

\author{S.~Damjanovic$^{2}$} for the NA60 Collaboration:
\author{R.~Arnaldi$^{10}$, 
R.~Averbeck$^{9}$, 
K.~Banicz$^{2,4}$, 
J.~Castor$^{3}$},
\author{
B.~Chaurand$^{7}$,
C.~Cical\`o$^{1}$
A.~Colla$^{10}$, 
P.~Cortese$^{10}$,}
\author{
S.~Damjanovic$^{2}$,
A.~David$^{2,5}$,
A.~De~Falco$^{1}$, 
A.~Devaux$^{3}$,} 
\author{A. Drees$^{9}$,
L.~Ducroux$^{6}$,
H.~En'yo$^{8}$, 
A.~Ferretti$^{10}$, 
M.~Floris$^{1}$,}  
\author{A.~F\"{o}rster$^{2}$,
P.~Force$^{3}$,
N.~Guettet$^{3}$,
A.~Guichard$^{6}$, 
H.~Gulkanian$^{11}$,}
\author{J.~Heuser$^{8}$,
M.~Keil$^{2,5}$,
L.~Kluberg$^{7}$, 
C.~Louren\c{c}o$^{2}$, 
J.~Lozano$^{5}$,}
\author{F.~Manso$^{3}$, 
A.~Masoni$^{1}$,
P.~Martins$^{2,5}$, 
A.~Neves$^{5}$, 
H.~Ohnishi$^{8}$,} 
\author{C.~Oppedisano$^{10}$,
P.~Parracho$^{2}$, 
P.~Pillot$^{6}$, 
G.~Puddu$^{1}$,}
\author{E.~Radermacher$^{2}$,
P.~Ramalhete$^{2}$, 
P.~Rosinsky$^{2}$, 
E.~Scomparin$^{10}$,}
\author{J.~Seixas$^{5}$,
S.~Serci$^{1}$, 
R.~Shahoyan$^{2,5}$, 
P.~Sonderegger$^{5}$,}
\author{H.J.~Specht$^{4}$,
R.~Tieulent$^{6}$, 
G.~Usai$^{1}$, 
R.~Veenhof$^{5,2}$, 
H.K.~W\"ohri$^{1}$}

\address{$^{1}$Univ.\ di Cagliari and INFN, Cagliari, Italy,
$^{2}$CERN, Geneva, Switzerland,
$^{3}$LPC, Univ.\ Blaise Pascal and CNRS-IN2P3, Clermont-Ferrand, France,
$^{4}$Univ.\ Heidelberg, Heidelberg, Germany,
$^{5}$IST-CFTP, Lisbon, Portugal,
$^{6}$IPN-Lyon, Univ.\ Claude Bernard Lyon-I and CNRS-IN2P3, Lyon, France,
$^{7}$LLR, Ecole Polytechnique and CNRS-IN2P3, Palaiseau, France,
$^{8}$RIKEN, Wako, Saitama, Japan,
$^{9}$SUNY Stony Brook, New York, USA,
$^{10}$Univ.\ di Torino and INFN, Italy,
$^{11}$YerPhI, Yerevan, Armenia}

\begin{abstract}
The NA60 experiment at the CERN SPS has studied low-mass muon pairs in
158 AGeV In-In collisions. A strong excess of pairs is observed above
the yield expected from neutral meson decays. After subtraction of the
decay sources, the shape of the resulting mass spectrum is largely
consistent with a dominant contribution from
$\pi^{+}\pi^{-}\rightarrow\rho\rightarrow\mu^{+}\mu^{-}$
annihilation. The associated $\rho$ spectral function exhibits
considerable broadening, but essentially no shift in mass. The
acceptance-corrected $p_{T}$ spectra have a shape atypical for radial
flow. They also significantly depend on mass, pointing to different
sources in different mass regions. Both mass and $p_{T}$ spectra are
compared to recent theoretical predictions.

\end{abstract}

\end{frontmatter}

\section{Introduction}
\label{intro}
Thermal dilepton production in the low-mass region is largely mediated
by the light vector mesons $\rho$, $\omega$ and $\phi$. Among these,
the $\rho$(770 MeV/c$^{2}$) is the most important, due to its strong
coupling to the $\pi\pi$ channel and its short lifetime of only 1.3
fm/c, much shorter than the lifetime of the fireball. Theoretical
predictions exist for changes both of the mass and the width which are
tied directly~\cite{Pisarski:mq,Brown:kk} or
indirectly~\cite{Rapp:1999ej} to chiral symmetry restoration. While
broadening of the $\rho$ is generally expected, changes of the mass
are controversial. For masses $>$0.9 GeV/c$^{2}$, other hadronic
processes set in, and partonic processes start to play a larger role
as well. The first NA60 results focused on the space-time averaged
spectral function of the $\rho$ \cite{Arnaldi:2006}; more details were
added recently~\cite{hq:2006}. The present paper therefore
concentrates on new developments: an analysis of the shape of the
excess mass spectra~\cite{hq:2006}, and in particular first results on
acceptance-corrected $p_{T}$ and $m_{T}$ spectra for different mass
windows. Mass and $p_{T}$ spectra are also compared to the latest
theoretical developments~\cite{rapphees:nn,rr:nn,zahed:nn}.

\section{Experimental results}
\label{sec:1}
\begin{figure*}[t!]
\centering
\includegraphics*[width=6.8cm, height=7.0cm,clip=, bb = 6 12 560 665]{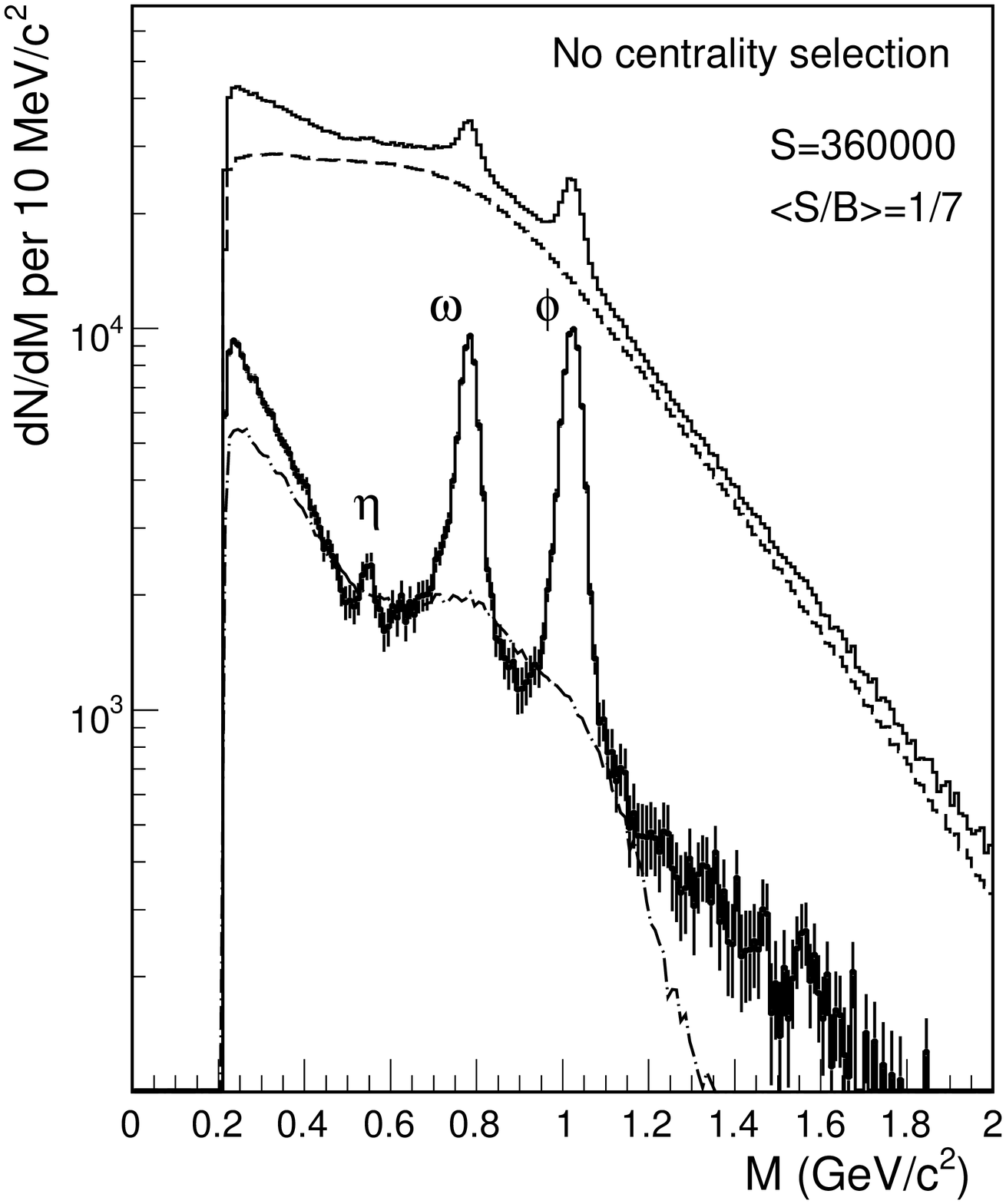}
\includegraphics*[width=6.8cm, height=7.0cm,clip=, bb= 0 12 560 665]{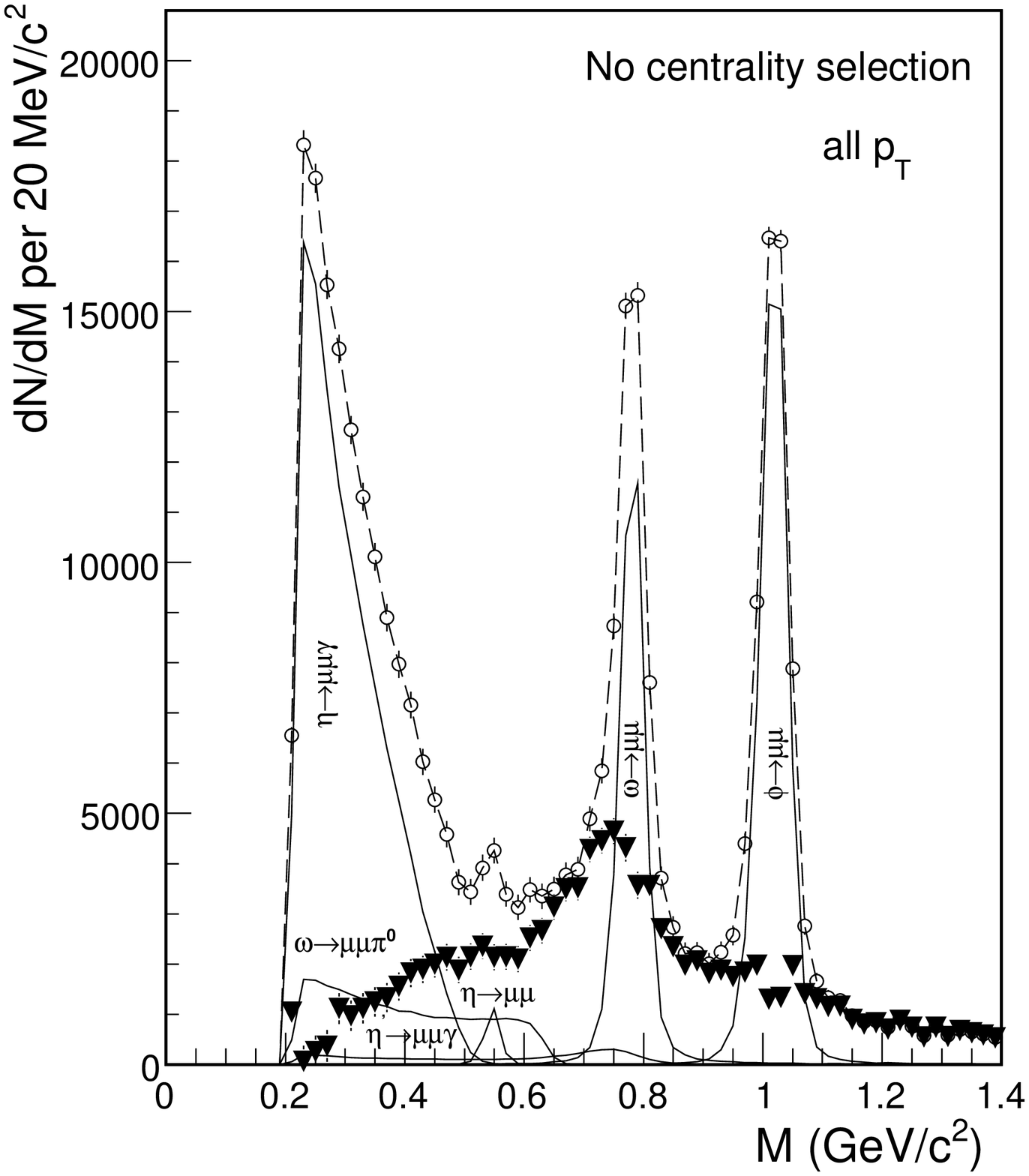}
\caption{Left: Mass spectra of the opposite-sign dimuons (upper
histogram), combinatorial background (dashed), signal fake matches
(dashed-dotted), and resulting signal (histogram with error
bars). Right:Isolation of an excess above the hadron decay cocktail
(see text). Total data (open circles), individual cocktail sources
(solid), difference data (thick triangles), sum of cocktail sources
and difference data (dashed).}
\label{fig1}
\end{figure*}

Details of the NA60 apparatus are contained
in~\cite{Gluca:2005,Keil:2005zq}, while the different analysis steps
(including the critical assessment of the combinatorial background
from $\pi$ and $K$ decays through event mixing) are described
in~\cite{Ruben:2005qm}. The results reported here were obtained from
the analysis of data taken in 2003 for In-In at 158 AGeV. The left
part of Fig.~\ref{fig1} shows the opposite-sign, background and signal
dimuon mass spectra, integrated over all collision centralities. After
subtracting the combinatorial background and the signal fake matches,
the resulting net spectrum contains about 360\,000 muon pairs in the
mass range 0-2 GeV/c$^2$, roughly 50\% of the total available
statistics. The associated average charged-particle multiplicity
density measured by the vertex tracker is $dN_{ch}/d\eta$ =120. The
vector mesons $\omega$ and $\phi$ are completely resolved; the mass
resolution at the $\omega$ is 20 MeV/c$^{2}$. A major part of the
subsequent analysis is done in four classes of collision centrality
(defined through the charged-particle multiplicity density):
peripheral (4$-$30), semiperipheral (30$-$110), semicentral (110$-$170) and
central (170$-$240).

The peripheral data can essentially be described by the expected
electromagnetic decays of the neutral mesons, i.e. the 2-body decays
of the $\eta$, $\rho$, $\omega$ and $\phi$ resonances and the Dalitz
decays of the $\eta$, $\eta^{'}$ and
$\omega$~\cite{Arnaldi:2006,hq:2006}. This is not true in the more
central bins, due to the existence of a strong excess. To isolate this
excess with {\it a priori unknown} characteristics without any fits,
we have used a novel procedure as shown in Fig.~\ref{fig1} (right),
made possible by the high data quality. For each centrality bin, the
cocktail of the decay sources is subtracted from the data by using
solely {\it local} criteria~\cite{Arnaldi:2006,hq:2006}. The $\rho,$
on the contrary, is not subtracted. Examples for the resulting
difference spectra are contained in Fig.~\ref{fig1} (right) for all
centralities and in Fig.~\ref{fig6} for the semicentral bin. The
qualitative features of the spectra are striking: a peaked structure
is always seen, residing on a broad continuum with a yield strongly
increasing with centrality, but remaining essentially centered around
the position of the nominal $\rho$ pole. Further details on the
subtraction procedure and the results, including a discussion of the
systematic errors (which reach up to 25\% in the continuum region),
can be found in~\cite{Arnaldi:2006,hq:2006}.

\begin{figure*}[t!]
\centering
\resizebox{1.0\textwidth}{!}{%
\includegraphics*[width=0.45\textwidth]{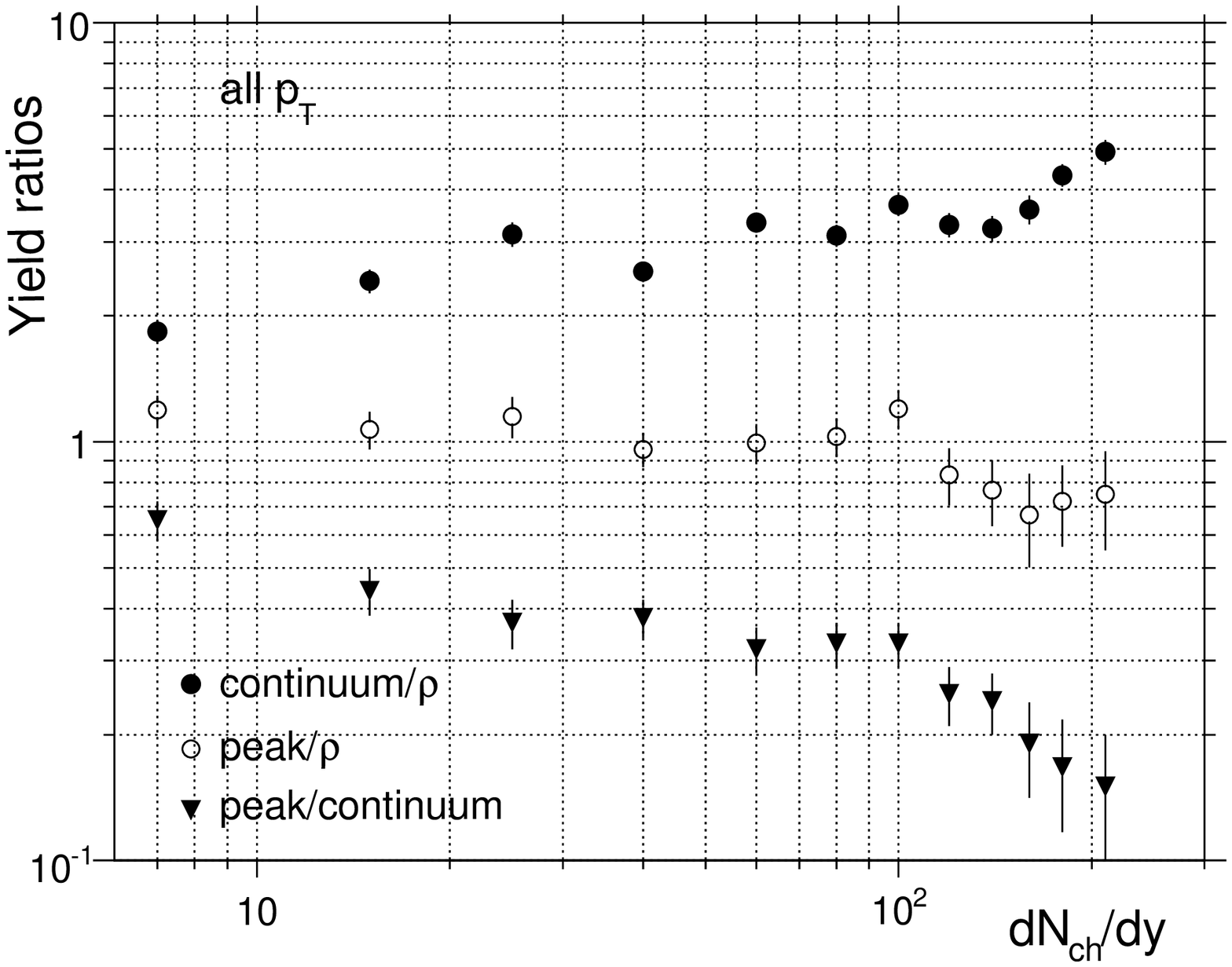}
\includegraphics*[width=0.45\textwidth]{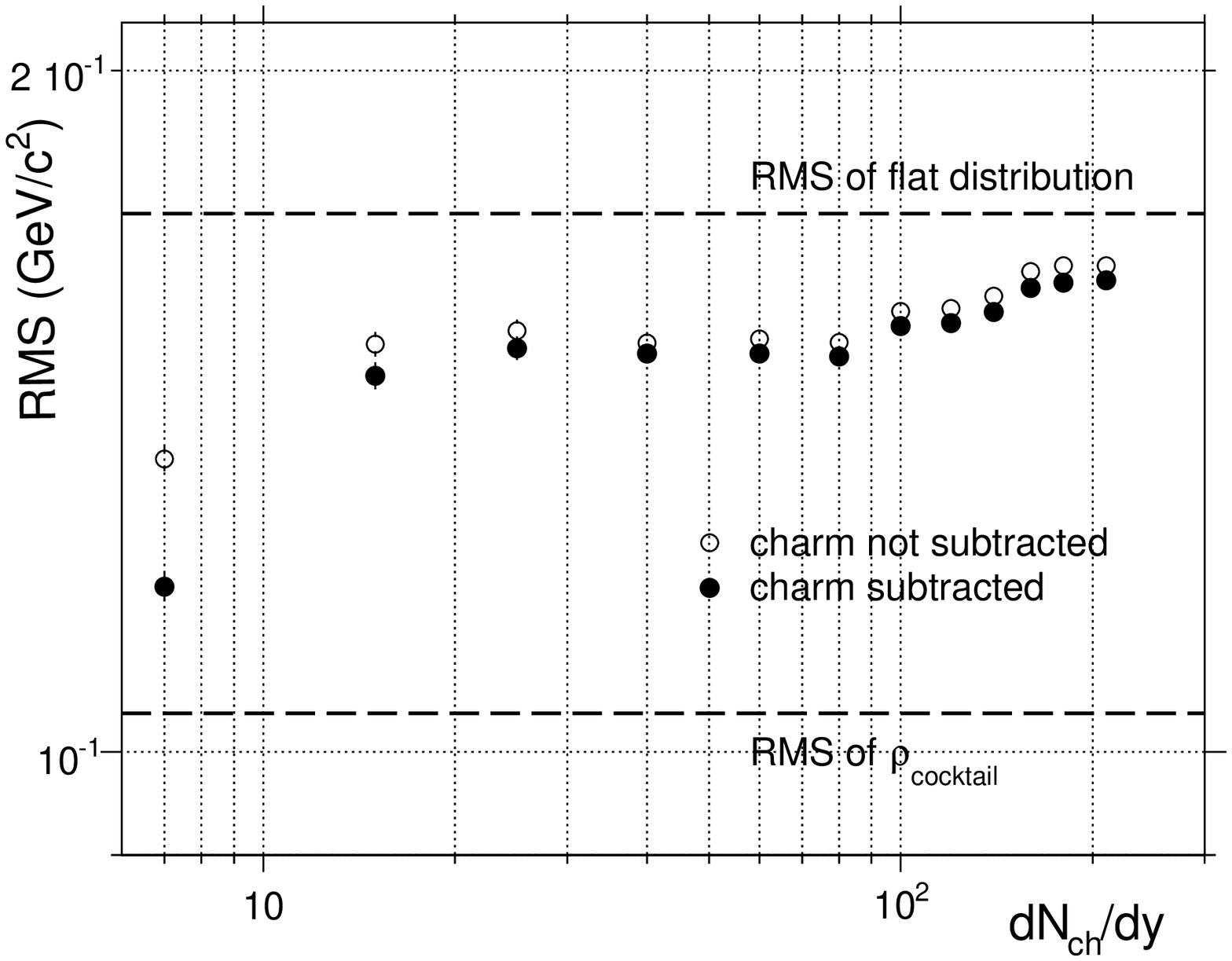}
}
\caption{Left: Yield ratios continuum/$\rho$, peak/$\rho$, and
peak/continuum (see text). The errors shown are purely statistical;
the point-to-point systematic errors are small compared to
them. Right: RMS of the excess mass spectra in the window
0.44$<$$M$$<$1.04 GeV/c$^{2}$. The subtraction of charm follows the
results from~\cite{Ruben:2005qm}, with the exception of the most
peripheral bin (100\% charm at $M$=1.1 GeV/c$^{2}$).}
\label{fig2}
\end{figure*}

A more quantitative analysis of the shape of the excess mass spectra
vs. centrality has recently been performed, using a finer subdivision
of the data into 12 centrality bins. The data (Fig.~\ref{fig1}, right)
were subdivided into 3 windows, one central (C) in the mass range
0.64$<$$M$$<$0.84 GeV/c$^{2}$, and two adjacent ones on the left (L)
and the right (R) with equal widths. The spectral shape can then
roughly be characterized by two yields: C-(L+R)/2 for the $\rho$-like
peak, and 3(L+R)/2 for the underlying continuum. The ratios
peak/$\rho$, continuum/$\rho$ and peak/continuum (where $\rho$ stands
for the cocktail $\rho$) are plotted in the left part of
Fig.~\ref{fig2}, compared to the RMS of the total mass interval
0.44$<$$M$$<$1.04
GeV/c$^{2}$ on the right. Three essential features should be
noted. First, the ratio peak/$\rho$ decreases from the most peripheral
to the most central bin by nearly a factor of 2, ruling out the naive
view that the shape can simply be explained by the cocktail $\rho$
residing on a broad continuum, independent of centrality. Second, the
sum of the ratios (peak+continuum)/$\rho$ is the total enhancement
factor relative to the cocktail $\rho$; it reaches about 5.5 in the
most central bin. Third, the centrality dependences are largely
consistent in the two plots. In particular, the flat part up to
dN$_{ch}$/dy=100 is followed by more rapid changes in all variables,
statistically highly significant for the ratio continuum/$\rho$ and
the RMS.

\begin{figure}[b!]
\centering
\includegraphics[width=0.45\textwidth]{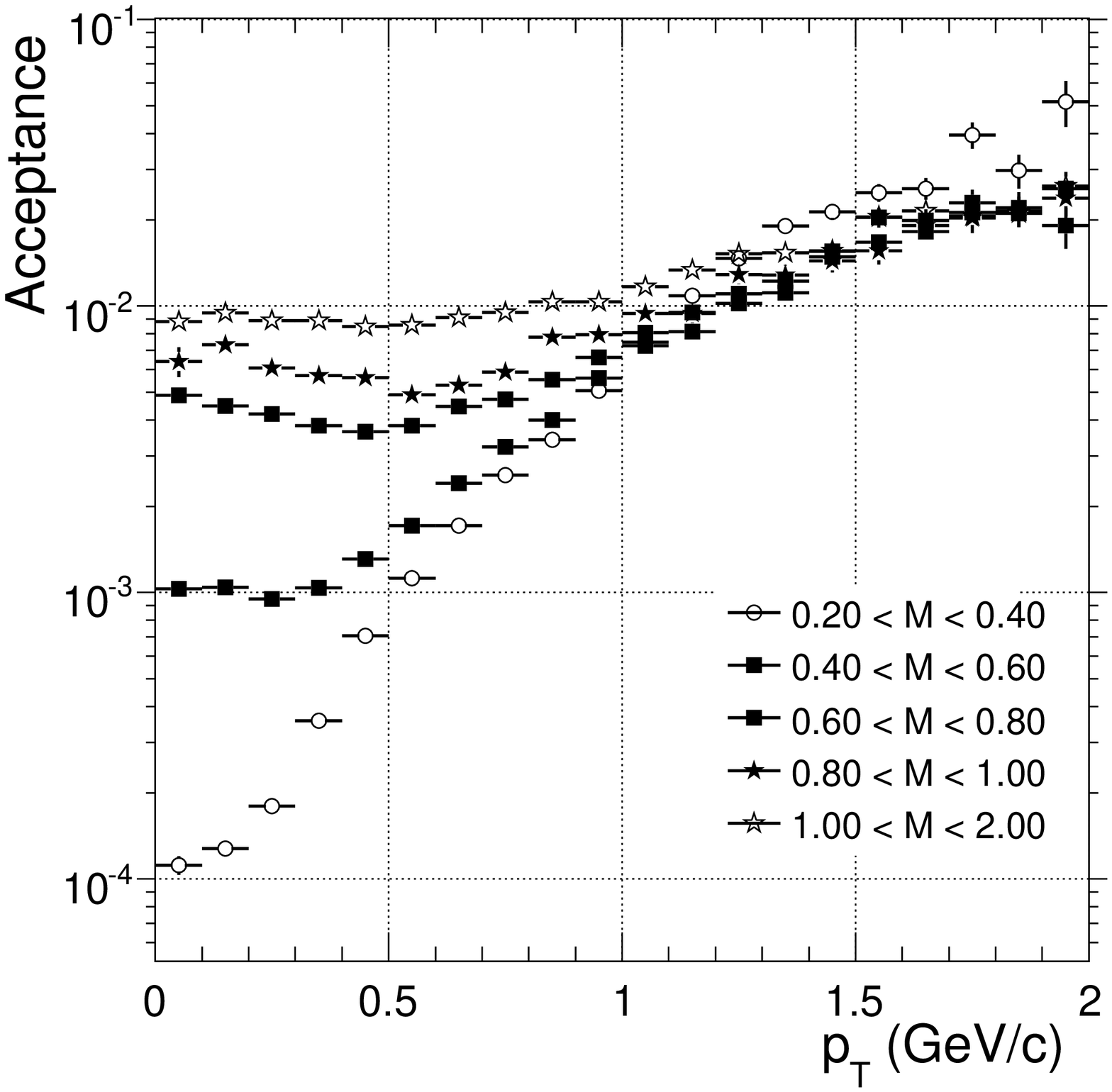} \hspace*{1cm}
\includegraphics[width=0.45\textwidth]{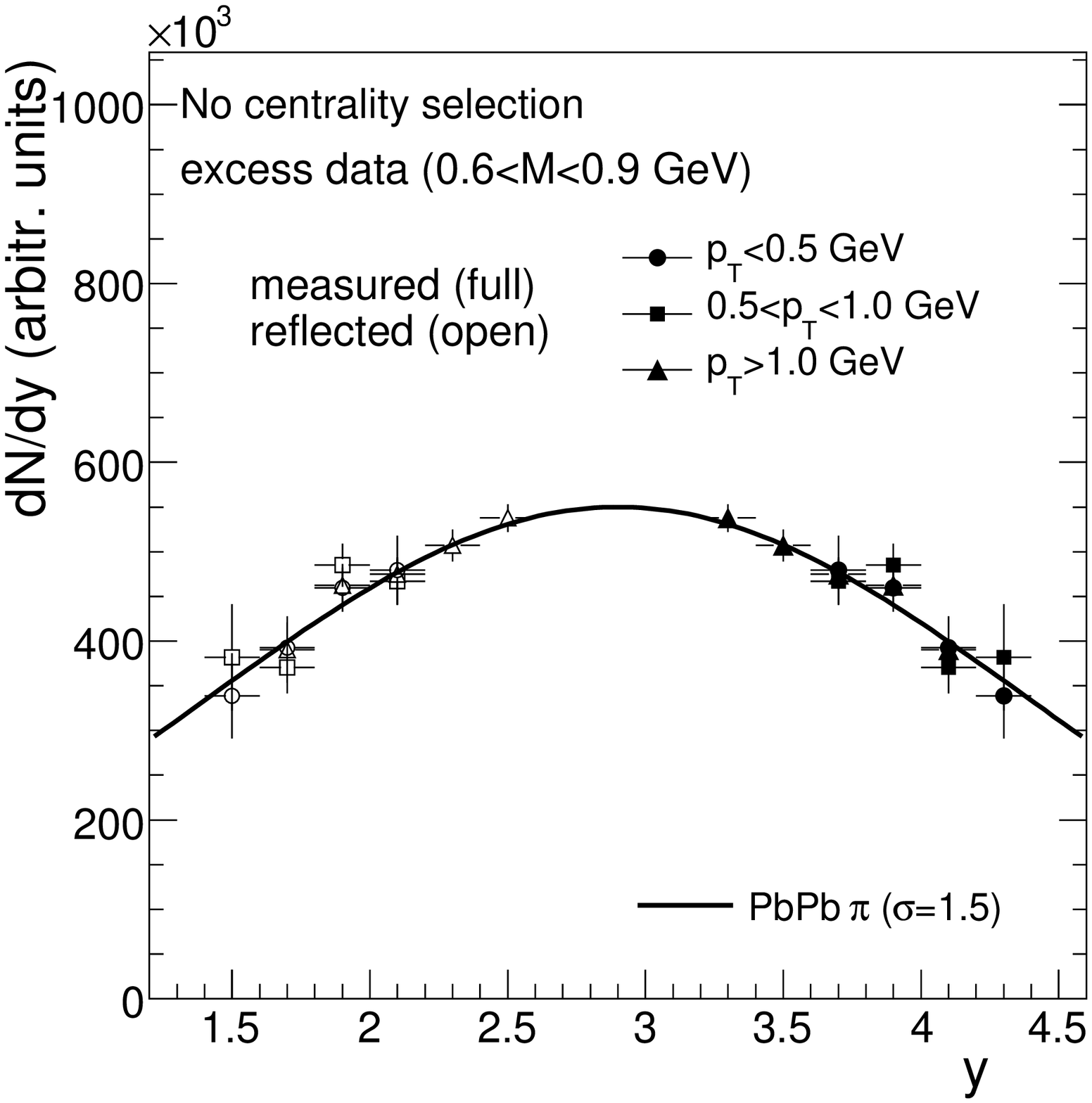}
\caption{Left: Acceptance relative to 4$\pi$ for different mass
windows. Right: Rapidity distribution of the excess data for the mass
window 0.6$<$$M$$<$0.9 GeV/$c^{2}$ and for three selected $p_{T}$
bins. The measured data (full markers) are reflected around
midrapidity (open markers).}
\label{fig3}
\end{figure}
Dependences on $p_T$ were so far shown only as mass spectra associated
with three different $p_T$ windows, without acceptance
correction~\cite{hq:2006}. We have now obtained acceptance-corrected
$p_T$ spectra associated with three different mass windows. The NA60
acceptance relative to $4\pi$ as a function of $p_T$ is shown in
Fig.~\ref{fig3} (left). In principle, the acceptance correction
requires a 3-dimensional grid in $M-p_T-y$ space. However, the low
populated bins in the phase space corners can introduce large errors
once the correction is applied. To overcome this problem, the
correction is performed in 2-dimensional $M-p_T$ space, using the
measured rapidity distribution as an input. The latter was determined
with an acceptance correction found, in an iterative way, from Monte
Carlo simulations matched to the data in mass and $p_T$. On the basis
of this rapidity distribution, 0.1 GeV/$c^2$ bins in mass and 0.2
GeV/$c$ bins in $p_T$ were used to determine the remaining
2-dimensional correction. Once corrected, the results were integrated
over the three extended mass windows 0.4$<$$M$$<$0.6, 0.6$<$$M$$<$0.9
and 1.0$<$$M$$<$1.4. GeV/c$^{2}$. The rapidity distribution of the
central mass window is shown in Fig.~\ref{fig3} (right) for three
different $p_T$ windows. It is interesting to note that it resembles
closely the distribution of inclusive pion production, as measured by
NA49 for Pb-Pb.

\begin{figure*}[h!]
\resizebox{1.0\textwidth}{!}{%
\includegraphics[width=0.62\textwidth]{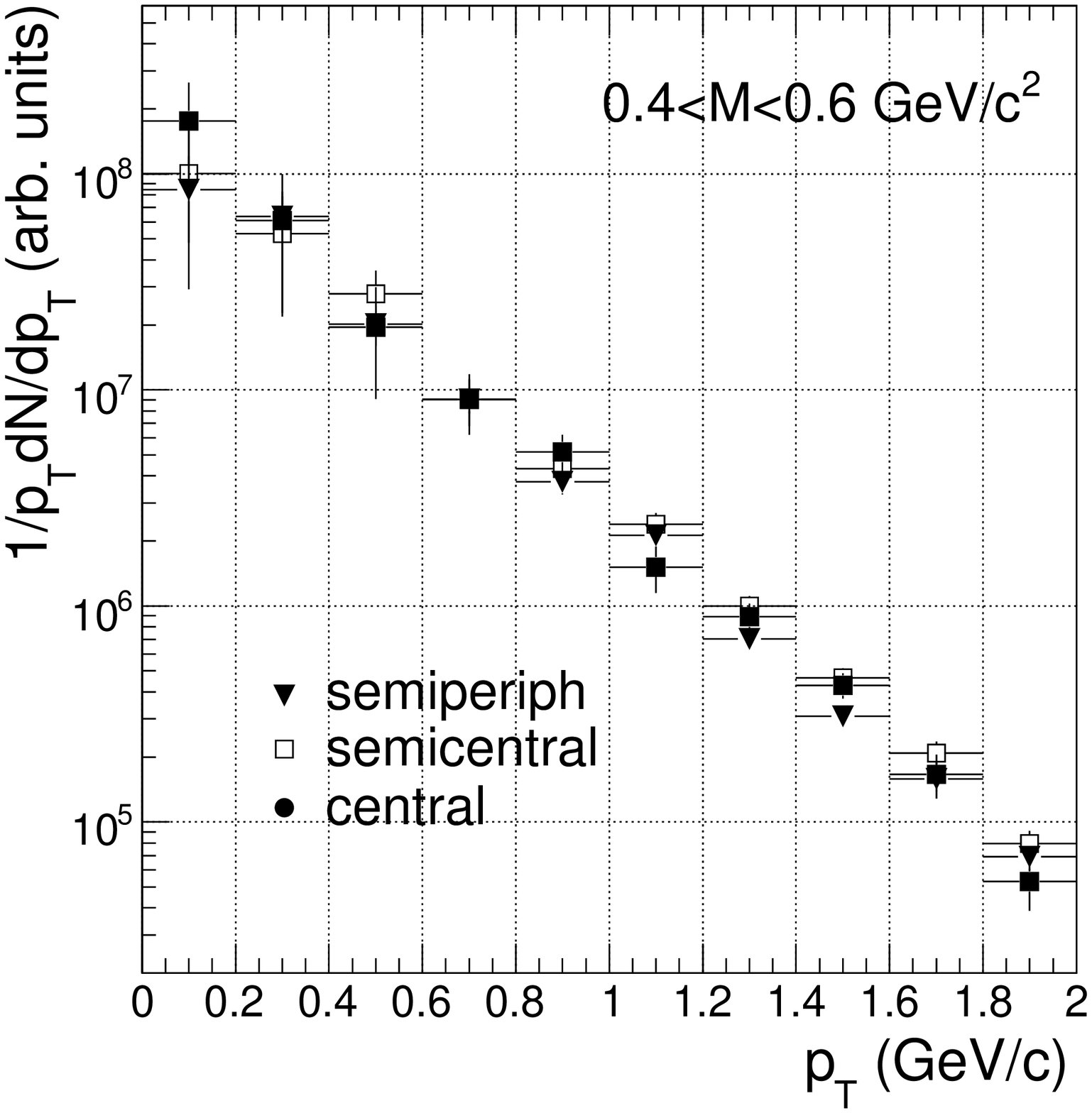}
\includegraphics[width=0.62\textwidth]{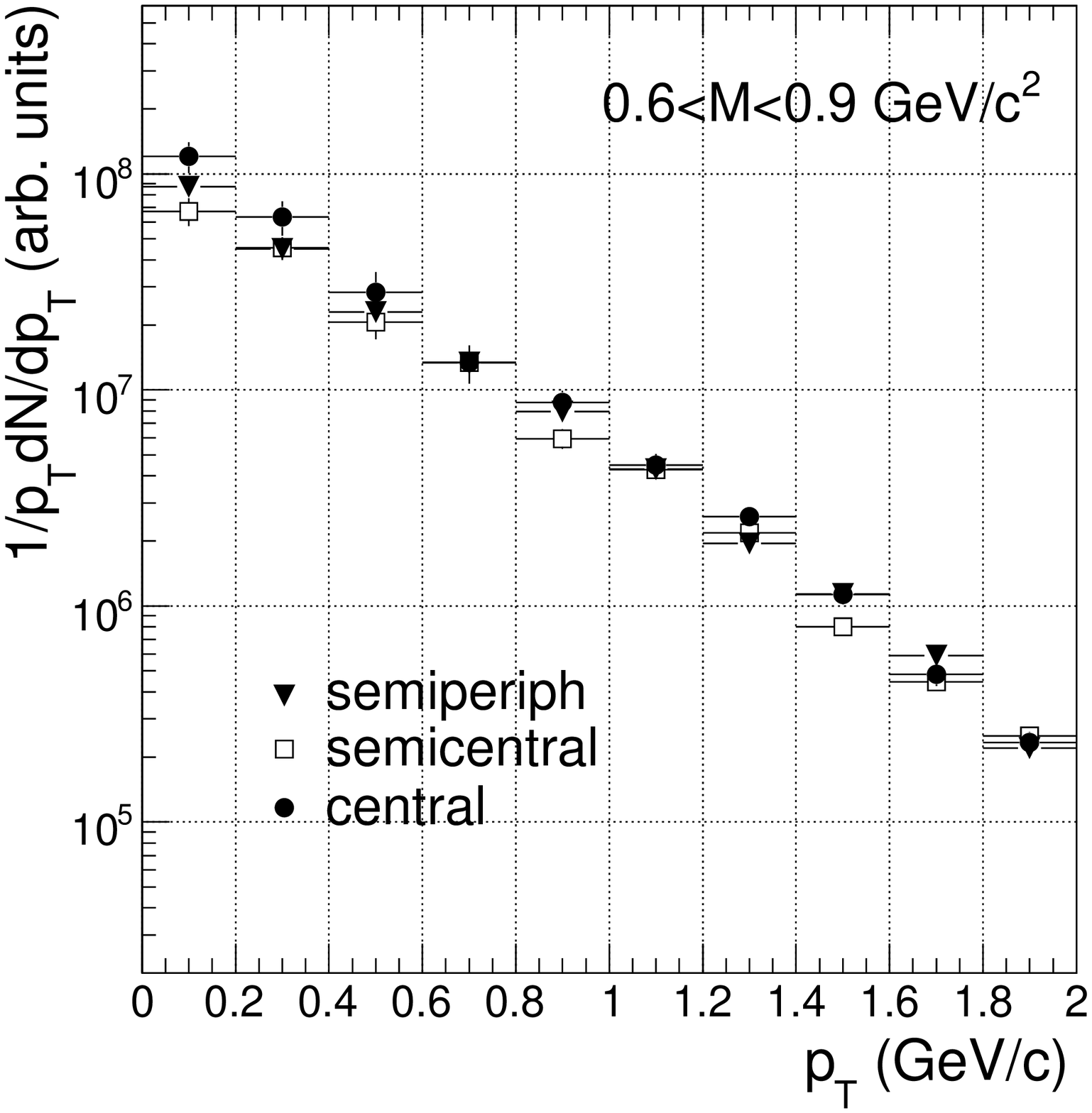}
\includegraphics[width=0.62\textwidth]{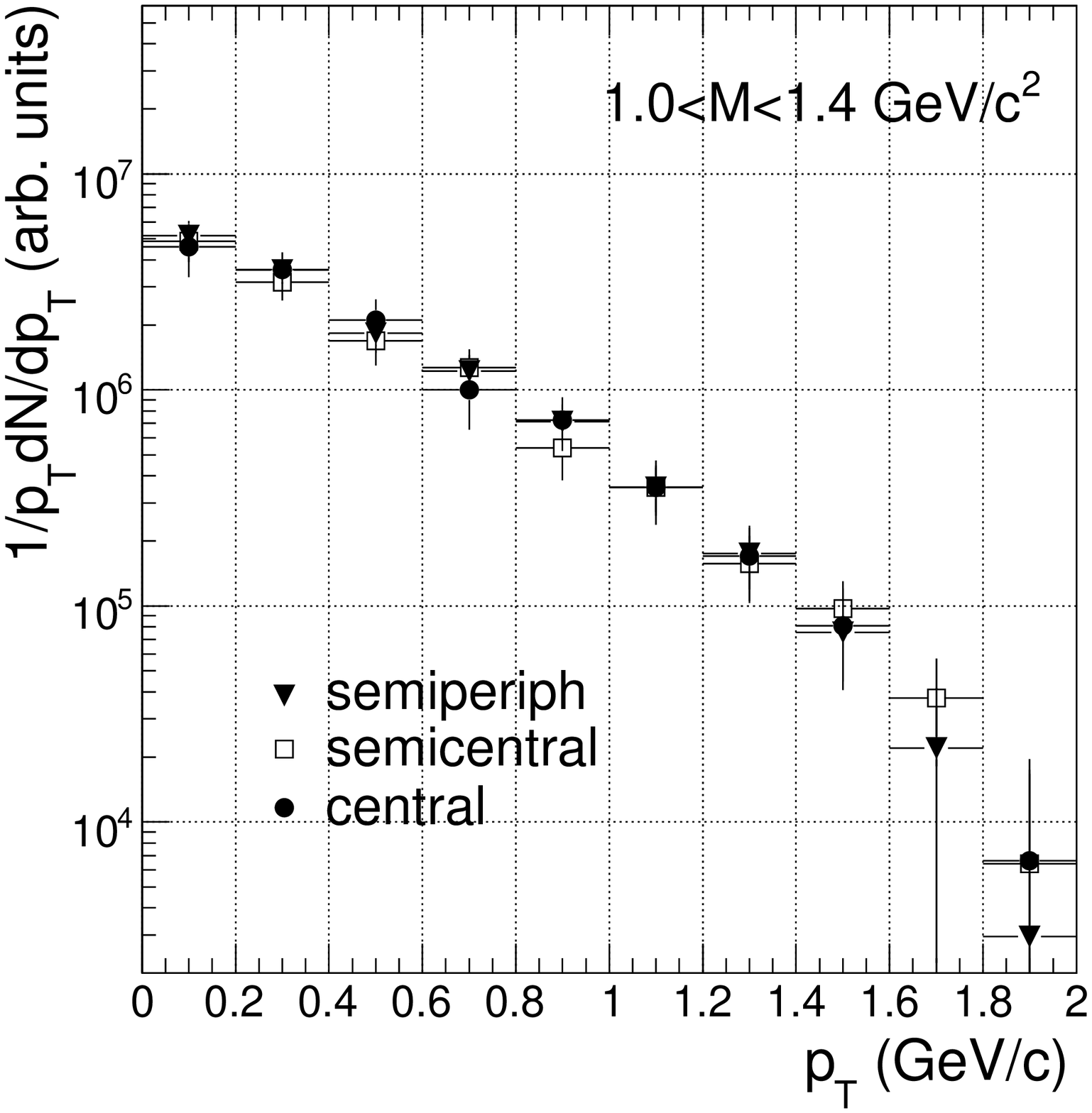}
}
\caption{Acceptance-corrected $p_{T}$ spectra for three mass windows
and for three centrality bins. For discussion of errors see text.}
\label{fig4}
\end{figure*}
\begin{figure*}[h!]
\centering
\includegraphics[width=0.45\textwidth]{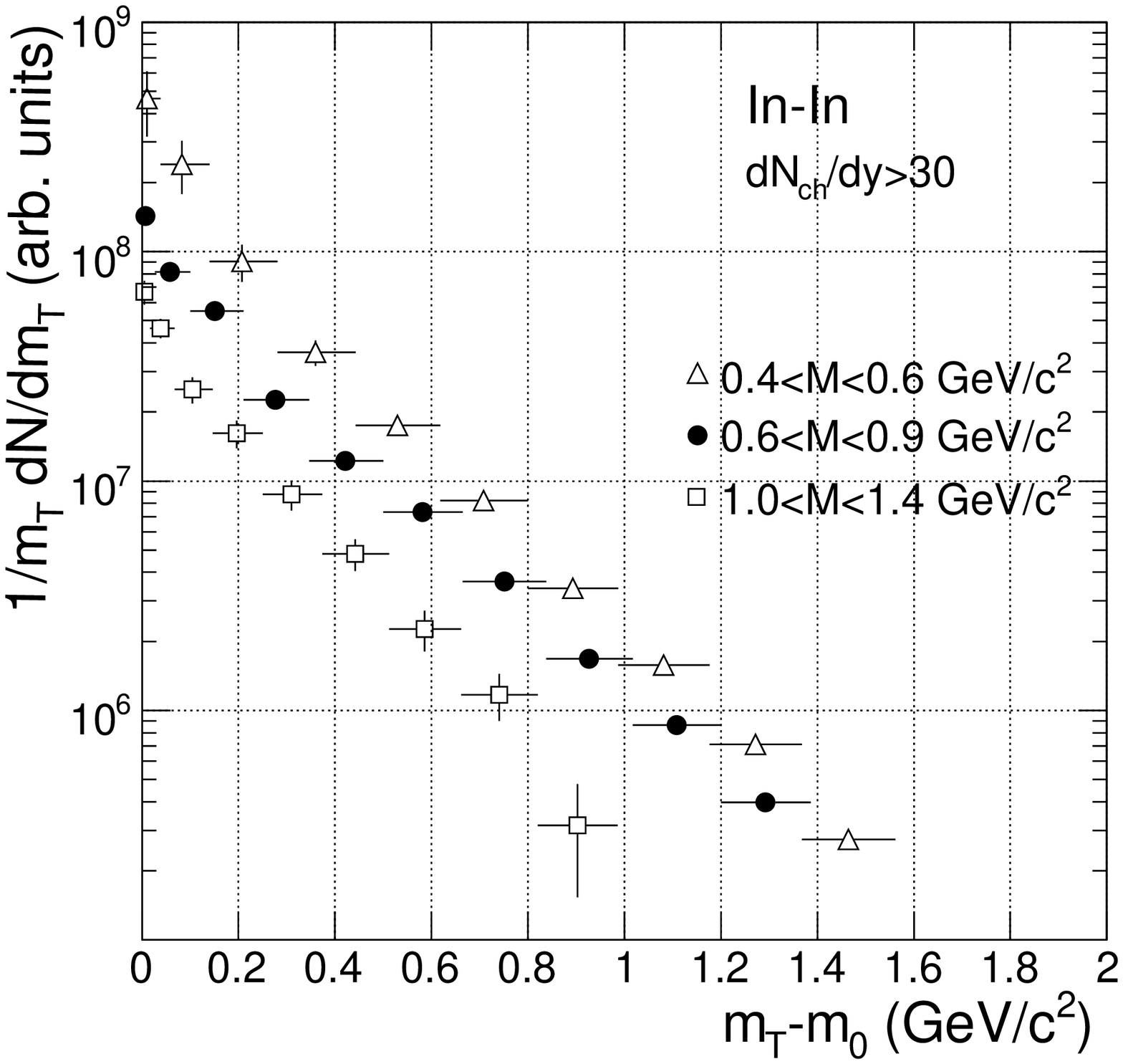}\hspace*{1cm}
\includegraphics[width=0.45\textwidth]{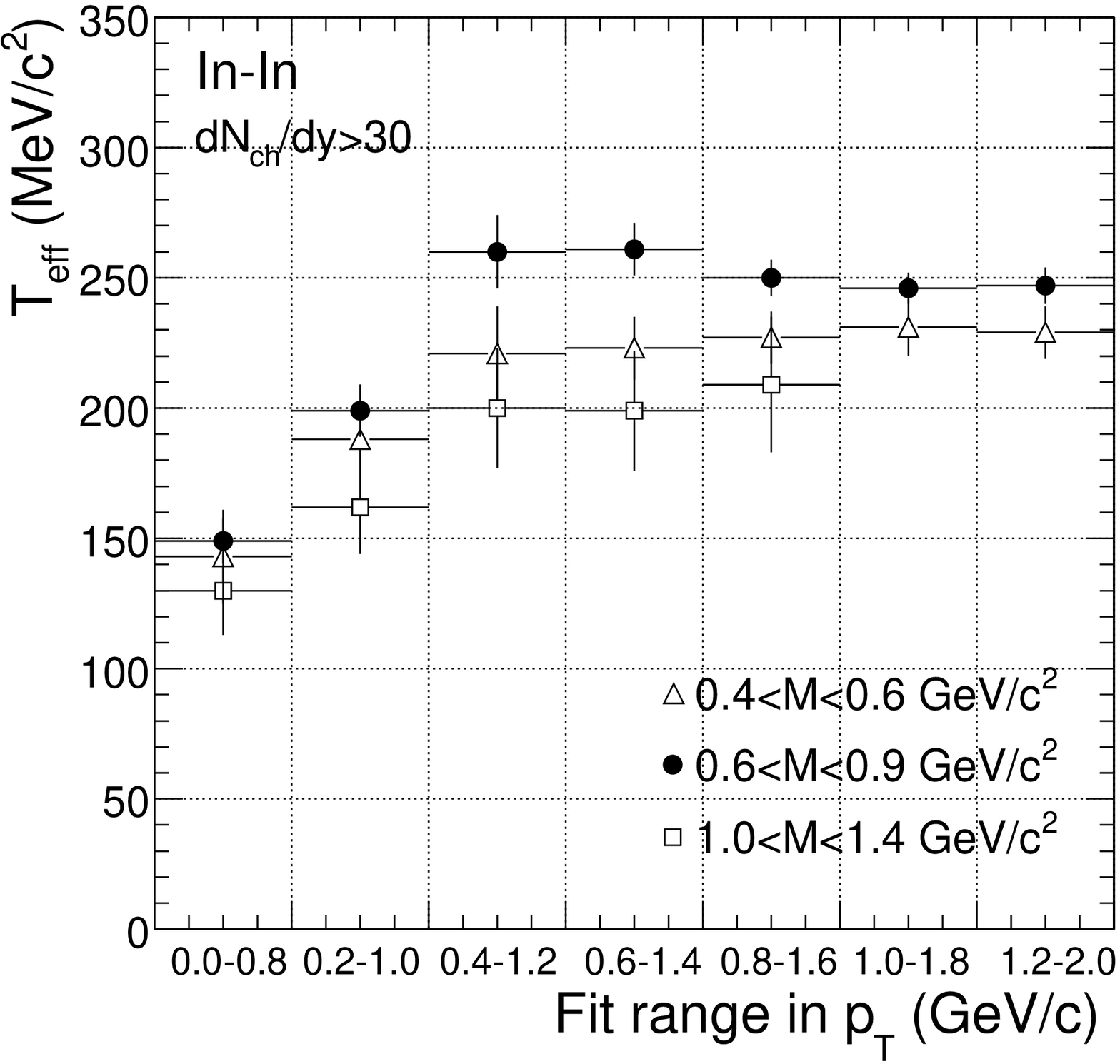}
\caption{Left: Acceptance-corrected $m_{T}$ spectra for three mass
windows, summed over centralities (excluding the peripheral
bin). Right: Inverse slope parameter $T_{eff}$ from differential fits
of the $m_{T}$ spectra (see text).}
\label{fig5}
\end{figure*}
%
The results for the acceptance-corrected $p_{T}$ spectra are
summarized in Fig.~\ref{fig4}. The errors shown are purely
statistical. Systematic errors arise from the acceptance corrections
including the rapidity distribution used, the subtraction of the
cocktail, and the subtraction of the combinatorial background plus
fake matches. For $p_{T}$$<$0.5 GeV/c, the combinatorial background
contributes most, ranging from 10 to 25\% for semiperipheral up to
central. For $p_{T}$$>$1 GeV/c, the statistical errors dominate. The
data show a significant dependence on mass, but hardly on
centrality. To bear out the differences in mass more clearly, the data
were summed over the three more central bins and plotted
vs. transverse mass $m_T$ in place of $p_T$, as shown in
Fig.~\ref{fig5} (left). The inverse slope parameter $T_{eff}$ as
determined from differential fits of the $m_{T}$ spectra with
$\exp(-m_{T}/T_{eff})$, using a sliding window in $p_{T}$, is plotted
on the right. The main features of the results are remarkable and
somewhat unexpected. At very low $m_{T}$, all spectra steepen rather
than flatten, as expected from radial flow, equivalent to very small
values of $T_{eff}$ (the $\phi$ resonance, placed just in between the
upper two mass windows, {\it does flatten} as expected). Moreover,
depending on the fit region, $T_{eff}$ covers an unusually large
dynamic range. Finally, the largest masses have the steepest $m_{T}$
spectrum, i.e. the smallest value of $T_{eff}$ everywhere, again
contrary to radial flow and to what is usually observed for
hadrons. All this suggests that different mass regions are coupled to
basically different emission sources.
%
\begin{figure*}[h!]
\centering
\includegraphics*[width=0.44\textwidth]{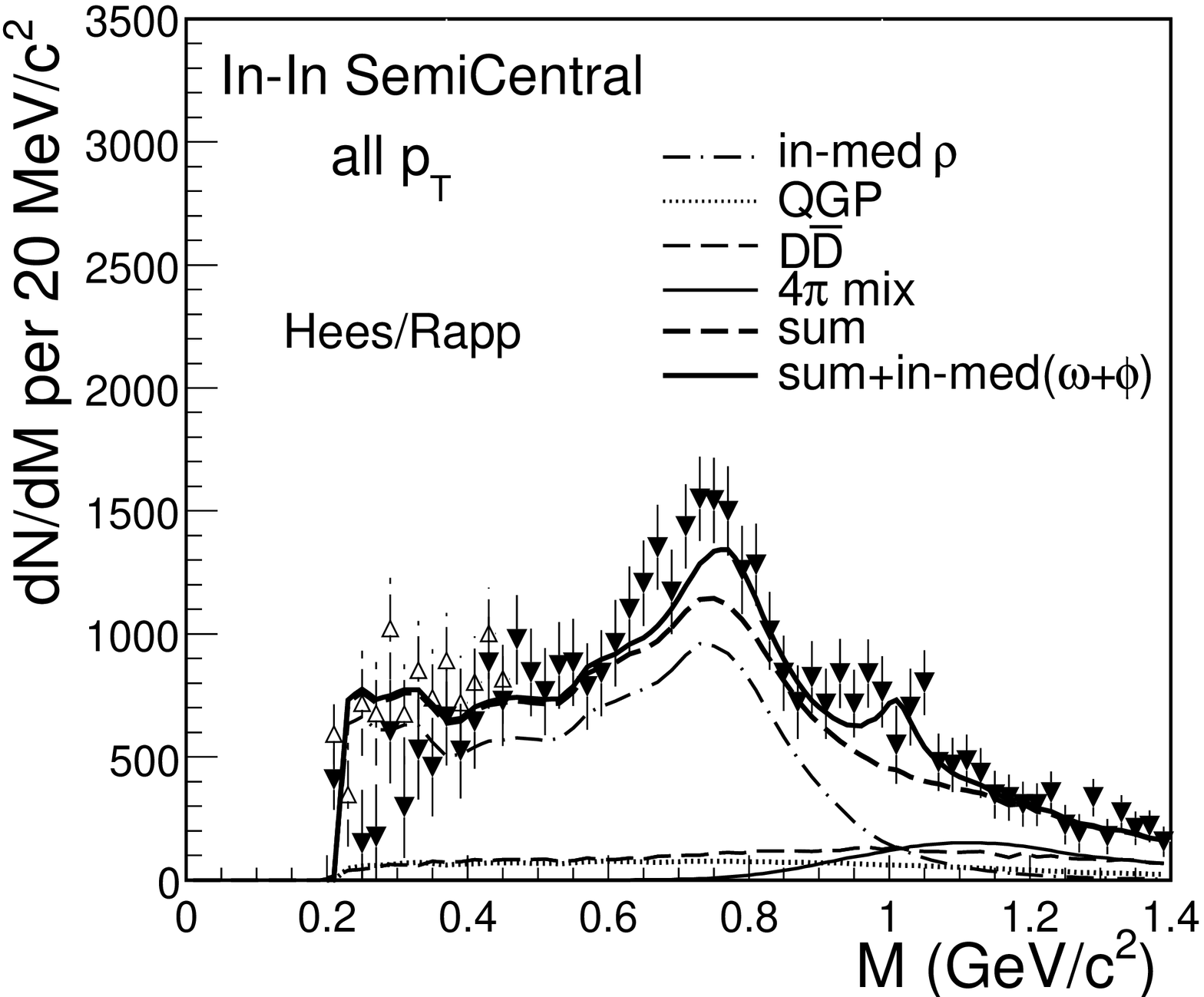}\hspace*{0.3cm}
\includegraphics*[width=0.44\textwidth]{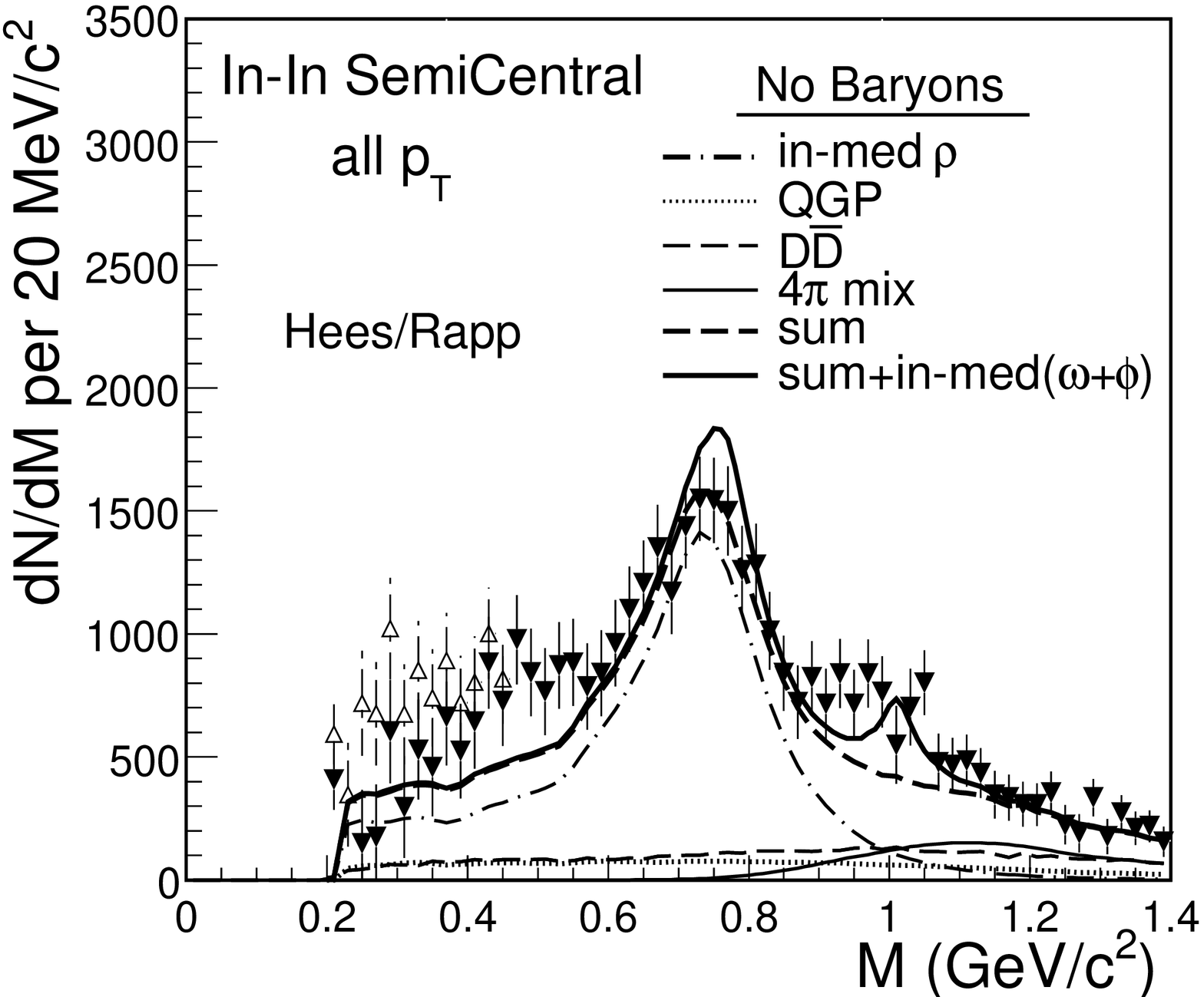}
\includegraphics*[width=0.44\textwidth]{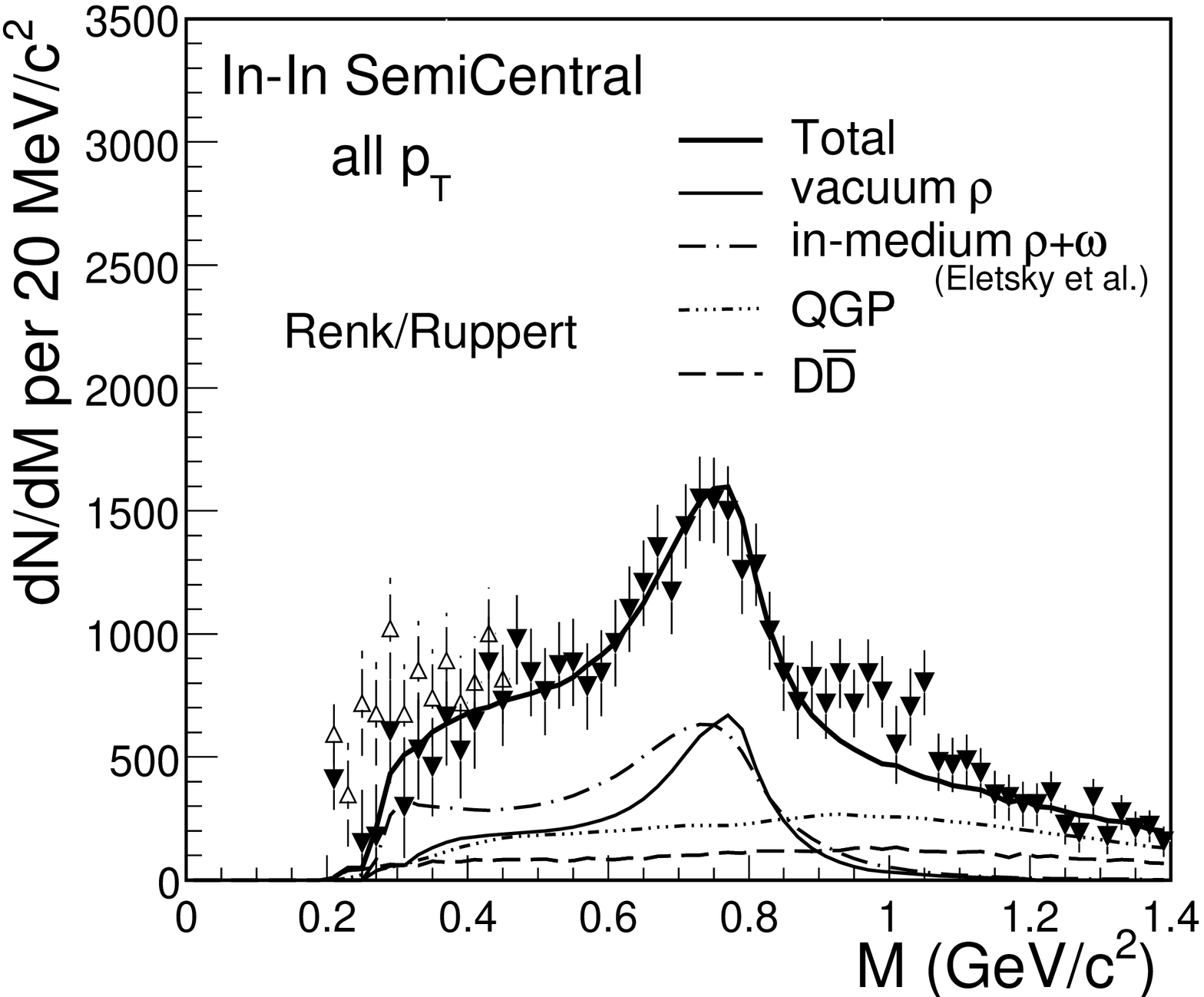}\hspace*{0.3cm}
\includegraphics*[width=0.44\textwidth]{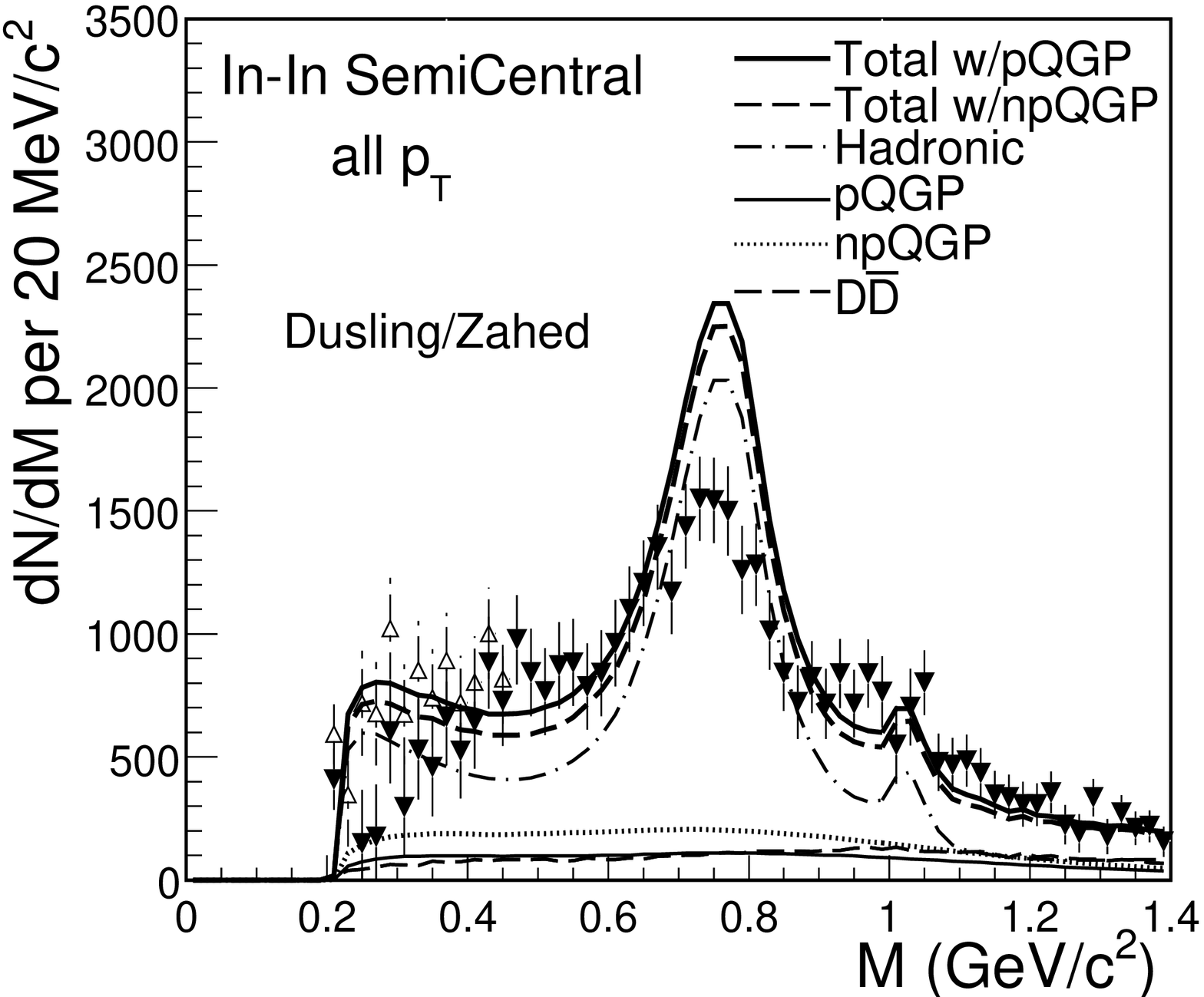}
\caption{Comparison of excess mass spectrum, with no $p_{T}$ cut, to
model predictions made for In-In at $dN_{ch}/d\eta$=140 (semicentral
bin)~\cite{rapphees:nn,rr:nn,zahed:nn}. Data errors are purely
statistical. The open data points show the difference spectrum
resulting from a decrease of the $\eta$ yield by 10\% (roughly the
high-$p_{T}$ limit of the measured $\eta$ yield).}
\label{fig6}
\end{figure*}
%
\section{Comparison to theoretical models}
\label{sec:3}
The qualitative features of the excess mass spectra shown above are
consistent with an interpretation as direct thermal radiation from the
fireball, dominated by
$\pi^{+}\pi^{-}\rightarrow\rho\rightarrow\mu^{+}\mu^{-}$ annihilation.
As explained in detail in \cite{Arnaldi:2006,hq:2006}, the excess mass
spectra, with {\it no} acceptance correction and no $p_T$ selection,
can be approximately interpreted as the spectral function of the
intermediate $\rho$, averaged over momenta and the complete space-time
history of the fireball.
\begin{figure*}[h!]
\centering
\includegraphics[width=0.43\textwidth]{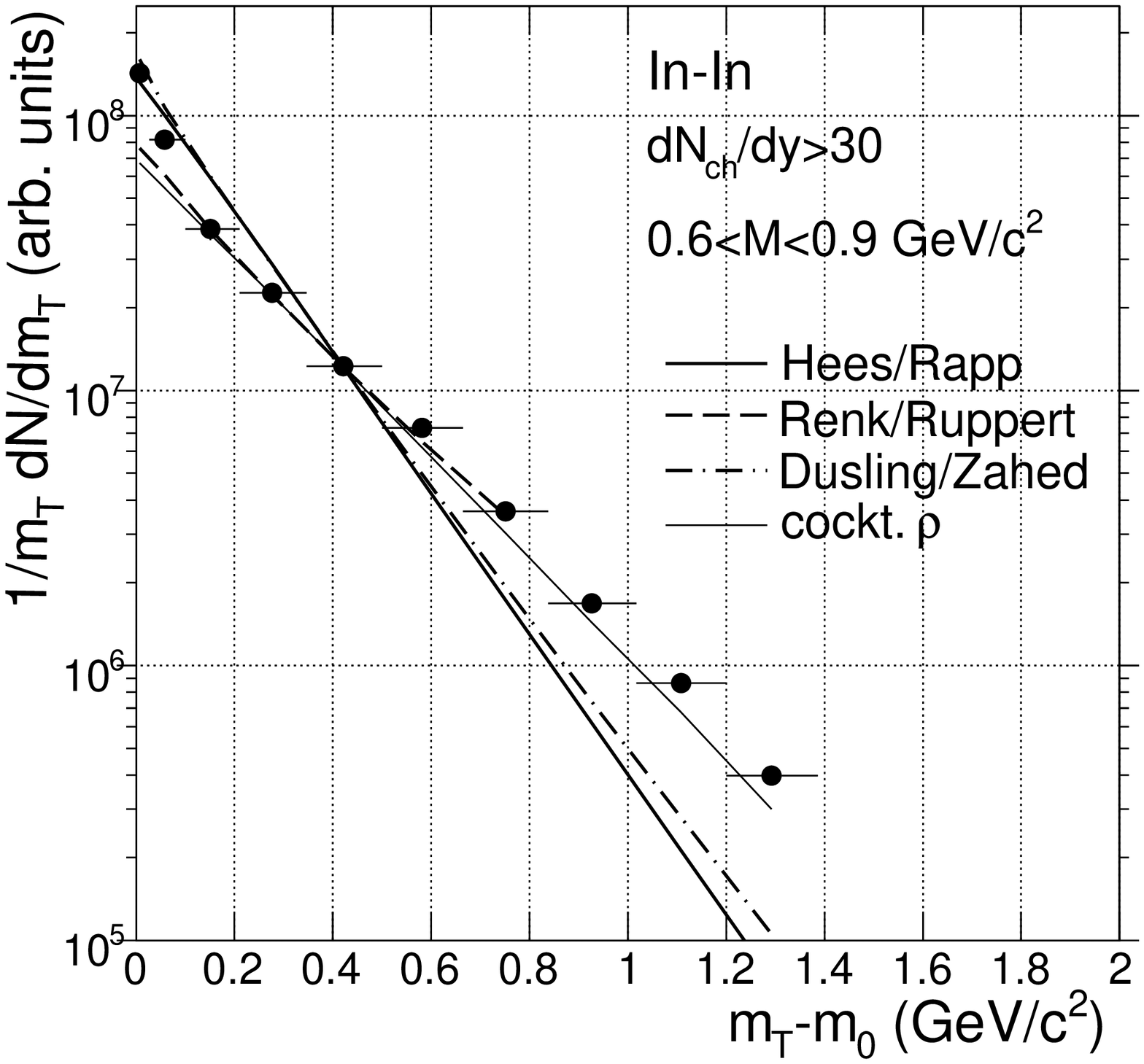}\hspace*{0.3cm}
\includegraphics[width=0.43\textwidth]{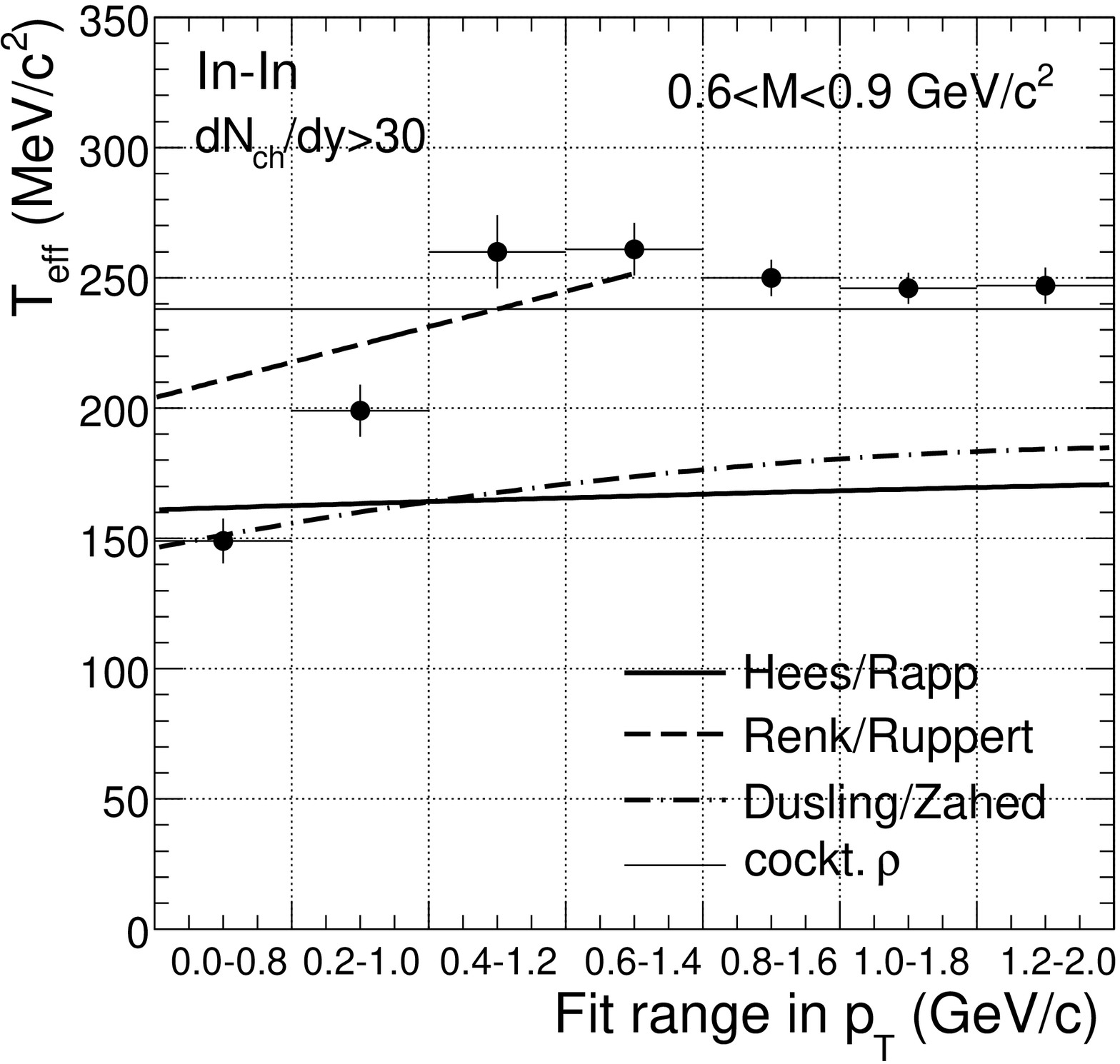}
\includegraphics[width=0.43\textwidth]{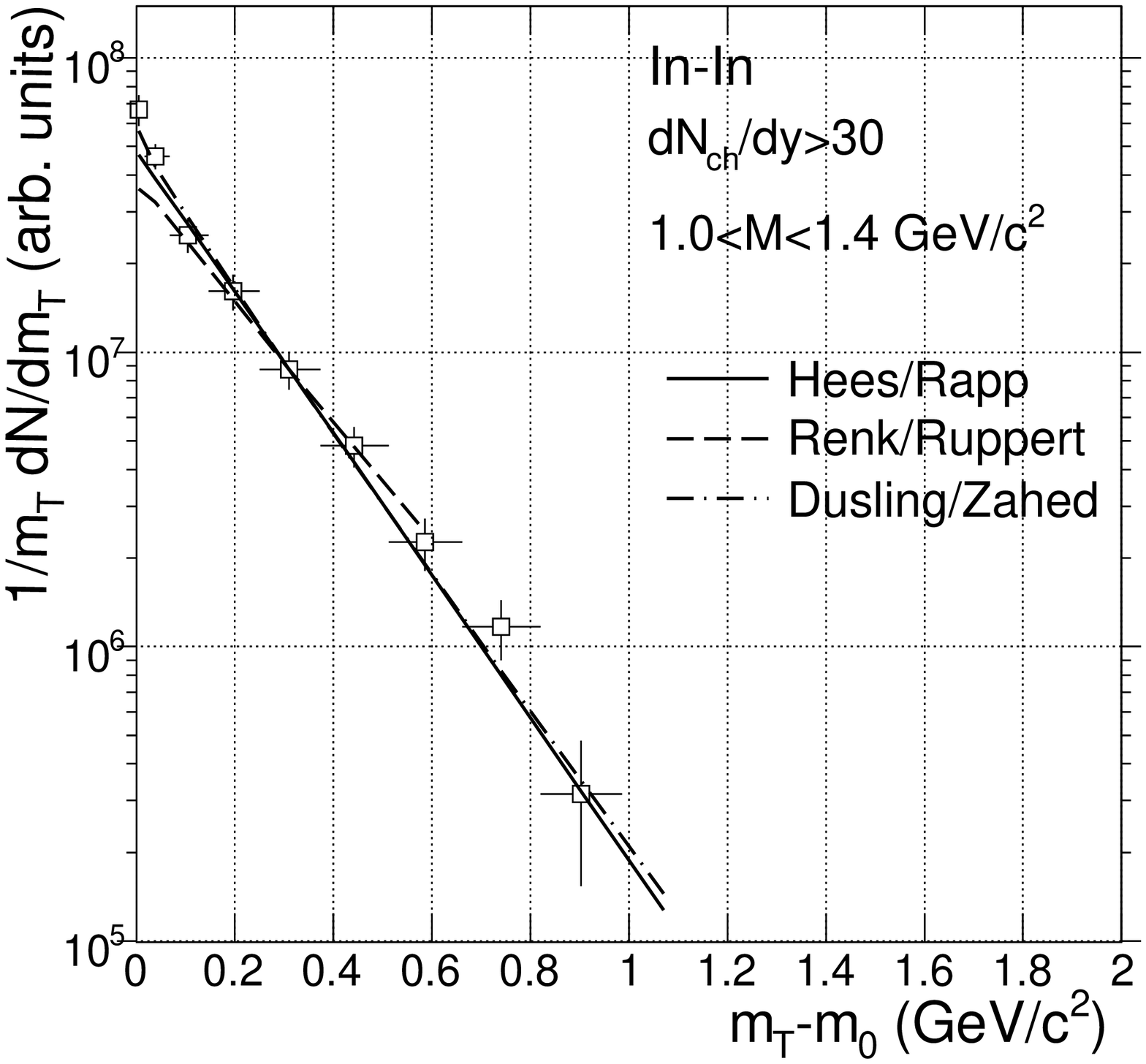}\hspace*{0.3cm}
\includegraphics[width=0.43\textwidth]{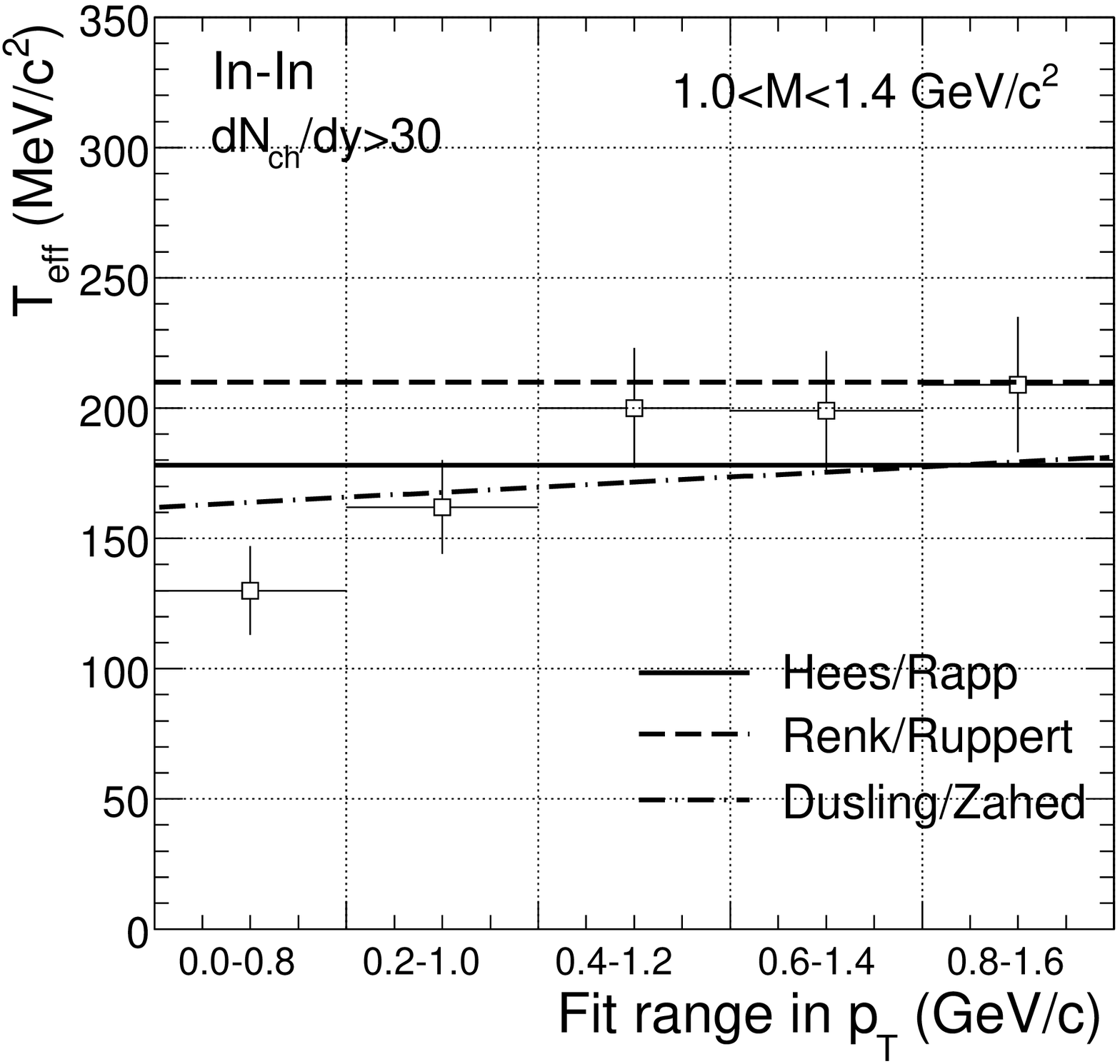}
\caption{Comparison of acceptance-corrected $m_{T}$ spectra (left) and
inverse slope parameters $T_{eff}$ from differential fits (right) to
model predictions made for In-In at
dN$_{ch}$/d$\eta$=140~\cite{rapphees:nn,rr:nn,zahed:nn} (the data
selection dN$_{ch}$/d$\eta$$>$30 corresponds to about the same value).
Mass windows 0.6$<$$M$$<$0.9 (upper) and 1.0$<$$M$$<$1.4 GeV/c$^{2}$
(lower).}
\label{fig7}
\end{figure*}
%
As already discussed in the first NA60 publications (see
\cite{Arnaldi:2006,hq:2006} and earlier ref. cited ibd.), the
moving-mass scenario \`a la Brown/Rho scaling~\cite{Brown:kk} was
found to be incompatible with the experimental excess mass
spectra. Even considering fireball variations within extremes, this
has basically remained unchanged. Other scenarios, characterized
essentially by a broadening of the $\rho$, have in the meantime been
refined~\cite{rapphees:nn} or newly developed~\cite{rr:nn,zahed:nn}.
A summary of the comparison of these latest theoretical results to the
excess mass spectrum for the semicentral bin is contained in
Fig.~\ref{fig6}. Hees/Rapp~\cite{rapphees:nn} have modified the
fireball part of the original predictions for In-In in the 2$\pi$
region (based on~\cite{Rapp:1999ej}) and have now also added 4$\pi$
processes, reflecting vector-axialvector mixing; the latter makes a
strong contribution for $M$$>$1 GeV/c$^{2}$. Renk/Ruppert~\cite{rr:nn}
use the spectral function of Eletsky et al. including baryonic effects
in the 2$\pi$ region, coupled to a strong partonic contribution which
dominates the yield for $M$$>$1
GeV/c$^{2}$. Dusling/Zahed~\cite{zahed:nn} use a chiral virial
approach including baryonic interactions, coupled again to some
partonic contribution which is on a similar level as the hadronic
yield for $M$$>$1 GeV/c$^{2}$. All models achieve a reasonable
description of the mass spectrum over the whole mass range, even in
absolute terms. For $M$$<$0.5 GeV/c$^{2}$, the decisive role played by
baryonic interactions in~\cite{rapphees:nn} is seen in the upper right
of Fig.~\ref{fig6} by switching them off. For $M$$>$1 GeV/c$^{2}$, the
agreement of all predictions with the data implies a low (if any)
sensitivity to the dynamics here, equivalent to hadron-parton
duality. The $m_T$ spectra predicted by the models above are compared
to the data in Fig.~\ref{fig7}. In contrast to the mass spectra, none
of the predictions is presently able to provide a satisfactory
description of the $m_{T}$ spectra over their full range. It seems
that the $m_{T}$ spectra are, in some sense, more sensitive to the
dynamics than the mass spectra. A quantitative understanding of the
$m_{T}$ spectra may thus help to discriminate between the different
sources and provide further insight beyond hadron-parton duality
(e.g. different radial flow effects for hadrons and partons).

Summarizing, the previously measured excess mass spectra and the new
accep\-tance-corrected $p_{T}$ and $m_{T}$ spectra present rather strict
boundary conditions to theoretical modeling. Beyond the $\rho$
spectral function itself, this may also lead to a better understanding
of the continuum part of the spectra.

\vspace*{-0.4cm}
%









\end{document}